\documentclass[fleqn,usenatbib]{mnras}

\usepackage[T2A,T1]{fontenc}
\usepackage[utf8]{inputenc}
\usepackage{ae,aecompl}

\usepackage{graphicx}	
\usepackage{amsmath}	
\usepackage{amssymb}	

\title[Jet in jet in M87]{Jet in jet in M87}

\author[D. N. Sob'yanin]{
Denis Nikolaevich Sob'yanin
\fontencoding{T2A}\selectfont
 (Денис Николаевич Собьянин)\thanks{E-mail: sobyanin@lpi.ru}
\fontencoding{T1}\selectfont
\\
I. E. Tamm Division of Theoretical Physics, P. N. Lebedev Physical Institute of the Russian Academy of Sciences,\\Leninskii Prospekt 53, Moscow 119991, Russia
\\
Moscow Institute of Physics and Technology (State University), Institutskii Pereulok 9, Dolgoprudnyi 141701, Russia
}

\date{Accepted 2017 July 7. Received 2017 July 7; in original form 2017 May 8}

\pubyear{2017}
\volume{471}
\pagerange{4121--4127}

\begin{document}
\label{firstpage}
\maketitle
\begin{abstract}
New high-resolution Very Long Baseline Interferometer observations of the prominent jet in the M87 radio galaxy show a persistent triple-ridge structure of the transverse 15-GHz profile with a previously unobserved ultra-narrow central ridge. This radio structure can reflect the intrinsic structure of the jet, so that the jet as a whole consists of two embedded coaxial jets. A relativistic magnetohydrodynamic model is considered in which an inner jet is placed inside a hollow outer jet and the electromagnetic fields, pressures and other physical quantities are found. The entire jet is connected to the central engine that plays the role of a unipolar inductor generating voltage between the jets and providing opposite electric currents, and the charge neutrality and current closure together with the electromagnetic fields between the jets can contribute to the jet stabilization. The constant voltage is responsible for the similar widening laws observed for the inner and outer jets. This jet-in-jet structure can indicate simultaneous operation of two different jet-launching mechanisms, one relating to the central supermassive black hole and the other to the surrounding accretion disc. An inferred magnetic field of $80$~G at the base is sufficient to provide the observed jet luminosity.
\end{abstract}



\section{Introduction}

M87 is a nearby dominant elliptical galaxy in the Virgo Cluster with the first discovered extragalactic jet \citep{Curtis1918}. It is one of the closest radio galaxies, with a distance of only about 16~Mpc \citep{BlakesleeEtal2009}. It is generally believed that the observed relativistic jet is powered by a supermassive black hole with a mass of $(3-6)\times10^9\text{ M}_\odot$ \citep{MacchettoEtal1997,GebhardtEtal2011,WalshEtal2013}, which corresponds to an active galactic nucleus (AGN). The proximity of M87 facilitates detailed investigation of the activity, which manifests itself throughout the spectrum from radio to very-high-energy $\gamma$-ray emission \citep{WilsonYang2002,PerlmanWilson2005,MadridEtal2007,AcciariEtal2009,BaesEtal2010,PerlmanEtal2011,HadaEtal2014}, and makes it a good candidate for broadening our knowledge of physical processes occurring in the central engine of the AGN.

An interest in theoretical studies of relativistic jets, arisen after the first pioneering works on the energy extraction from a black hole and the jet origin \citep{Penrose1969,Blandford1976,Lovelace1976,BlandfordZnajek1977,BlandfordPayne1982}, is yet more growing nowadays because of the necessity to firmly establish the nature of the jet and the exact mechanism of its launching, collimation, stabilization and propagation in the external medium \citep{ChiuehEtal1991,ApplCamenzind1993,IstominPariev1996,Fendt1997,LyndenBell2003,BeskinNokhrina2009,porthFendt2010,PorthEtal2011,Cao2012,ColgateEtal2015,TchekhovskoyBromberg2016,FengEtal2016,YangReynolds2016,EnglishEtal2016,BritzenEtal2017,SobacchiEtal2017}. For the same reason, a significant role is given to high-resolution observational research \citep{JunorEtal1999,KovalevEtal2007,HadaEtal2011,HadaEtal2012,HadaEtal2014,DoelemanEtal2012,AkiyamaEtal2015}, which can clarify some key features of the internal jet structure. The brightness and closeness of the M87 jet allowed one to reach an unprecedented ultra-high resolution down to $50\text{ }\mu$arcsec in Very Long Baseline Interferometer (VLBI) radio observations, which corresponds to only $6-10$ Schwarzschild radii \citep{HadaEtal2016,KimM87Etal2016}.

All previous studies showed that the M87 jet is almost parabolic at relatively small ($<10^5$ Schwarzschild radii) distances from the base \citep{AsadaNakamura2012}, characterized by a limb-brightened transverse profile of radio intensity at various frequencies, and then can be considered e.g. in the model of a magnetohydrodynamic nozzle \citep{NakamuraAsada2013}. However, new radio observations of M87 at 15~GHz (2~cm) using the NRAO Very Long Baseline Array in concert with the phased Y27 Very Large Array clearly indicate the existence of a persistent triple-ridge structure across the jet \citep{Hada2017}. Moreover, detection of an ultra-narrow central radio ridge in these observations sets up problems in explaining the effect with the standard spine-sheath jet model usually used when explaining the appearance of the limb brightness \citep{GhiselliniEtal2005} and poses a question whether we really observe a single jet with some decaying radial velocity profile. In this paper, I study the possibility that observing the above structure in the radio image, we can in fact deal with a pure jet-in-jet structure in M87: the inner jet is placed inside the outer annular jet. This circumstance can be evidence of simultaneous operation of two different jet-launching mechanisms, one relating to the black hole and the other to the accretion disc.

\section{Jet in jet}

The whole jet is governed by the Maxwell equations
\begin{gather}
\label{divE}
\operatorname{div}\mathbf{E}=4\pi\rho_e,
\\
\label{divB}
\operatorname{div}\mathbf{B}=0,
\\
\label{rotE}
\operatorname{curl}\mathbf{E}=-\frac{\partial\mathbf{B}}{\partial t},
\\
\label{rotB}
\operatorname{curl}\mathbf{B}=4\pi\mathbf{j}+\frac{\partial\mathbf{E}}{\partial t},
\end{gather}
the condition of infinite conductivity
\begin{equation}
\label{forceFree}
\mathbf{E}=-\mathbf{v}\times\mathbf{B},
\end{equation}
and the laws of conservation of matter
\begin{equation}
\label{matterConservation}
\frac{\partial\gamma\rho}{\partial t}+\operatorname{div}\gamma\rho\mathbf{v}=0,
\end{equation}
energy
\begin{equation}
\label{energyConservation}
\frac{\partial}{\partial t}\biggl(\gamma^2 \rho h-p+\frac{E^2+B^2}{8\pi}\biggr)+\operatorname{div}\biggl(\gamma^2 \rho h\mathbf{v}+\frac{\mathbf{E}\times\mathbf{B}}{4\pi}\biggr)=0,
\end{equation}
and momentum
\begin{align}
\label{momentumConservation}
&\frac{\partial}{\partial t}\biggl(\gamma^2 \rho h\mathbf{v}+\frac{\mathbf{E}\times\mathbf{B}}{4\pi}\biggr)\nonumber\\
&+\operatorname{div}\biggl[\biggl(p+\frac{E^2+B^2}{8\pi}\biggr)\mathbf{I}+\gamma^2 \rho h\mathbf{v}\mathbf{v}-\frac{\mathbf{E}\mathbf{E}+\mathbf{B}\mathbf{B}}{4\pi}\biggr]=0.
\end{align}
In the above equations, we put the speed of light $c=1$, and the electric and magnetic fields are denoted by $\mathbf{E}$ and $\mathbf{B}$, respectively, $\rho_e$ and $\mathbf{j}$ are the volume charge and current densities, respectively, $\mathbf{v}$ is the velocity of the jet plasma at a given point, $\gamma=(1-v^2)^{-1/2}$ is the corresponding Lorentz factor, $h=1+\varepsilon+p/\rho$ is the specific relativistic enthalpy, $\varepsilon$ is the specific internal energy, $p$ is the pressure and $\rho$ is the mass density in the comoving reference frame. We also use notations $\mathbf{a}\mathbf{b}=||a_ib_j||$ for a dyad, $\mathbf{I}=||\delta_{ij}||$ for the unit tensor, and $\operatorname{div}\mathbf{T}=\nabla\cdot\mathbf{T}=||\partial T_{ji}/\partial x_j||$ for the divergence of a tensor $\mathbf{T}=||T_{ij}||$. The system is complemented by an equation of state $p=p(\rho,\varepsilon)$.

We will consider the case of stationarity and pure cylindrical symmetry. We may write in the cylindrical coordinates the velocity
\begin{equation}
\label{velocity}
\mathbf{v}=v_\phi\mathbf{e}_\phi+v_z\mathbf{e}_z
\end{equation}
and electromagnetic fields
\begin{gather}
\label{electricField}
\mathbf{E}=E_r\mathbf{e}_r=(v_z B_\phi-v_\phi B_z)\,\mathbf{e}_r,
\\
\label{magneticField}
\mathbf{B}=B_\phi\mathbf{e}_\phi+B_z\mathbf{e}_z.
\end{gather}
The components $v_\phi$, $v_z$, $B_\phi$, $B_z$ as well as the other scalar quantities depend on $r$ only, the distance from the symmetry axis to a given point. Equations \eqref{velocity}--\eqref{magneticField} mean that the lines of the matter flow and the magnetic field lines are helices lying on a cylindrical tube, and the electric field lines are orthogonal to the lateral tube surface and radially diverge from (or converge to) the symmetry axis. In the stationary case, equations \eqref{divB}, \eqref{rotE} and \eqref{forceFree}--\eqref{energyConservation} are then automatically satisfied, while equations \eqref{divE} and \eqref{rotB} simply determine the charge and current densities
\begin{gather}
\label{charge}
\rho_e=\frac{(r E_r)'}{4\pi r},
\\
\label{current}
\mathbf{j}=-\frac{B'_z}{4\pi}\,\mathbf{e}_\phi+\frac{(r B_\phi)'}{4\pi r}\,\mathbf{e}_z,
\end{gather}
where prime denotes the $r$ derivative. Consequently, the momentum conservation law \eqref{momentumConservation}, which now takes the form
\begin{equation}
\label{momentumConservation2}
\biggl(p+\frac{B_z^2+B_\phi^2-E_r^2}{8\pi}\biggr)'+\frac{B_\phi^2-E_r^2}{4\pi r}-\gamma^2 \rho h\frac{v_\phi^2}{r}=0,
\end{equation}
plays the major role in determining the jet equilibrium.

An equivalent approach to stationary axisymmetric plasma systems is based on the Grad-Shafranov equation \citep{Fendt1997,BeskinNokhrina2009} and earlier allowed one to study jet collimation with the help of a steady-state trans-field force-balance equation \citep{ChiuehEtal1991,ApplCamenzind1993}, consider differential rotation in the force-free approximation \citep{Fendt1997}, and then extend the results to the full time-dependent two-dimensional solution for the collimated jet structure close to the disc and central object \citep{porthFendt2010,PorthEtal2011}. The equivalence of the two approaches is evident as in the stationary axisymmetric case, when the radius of a magnetic tube may change with the distance from the jet base, we have the same integrals of motion as in the Grad-Shafranov approach that follow from conservation of the magnetic flux in the tube and, respectively, matter conservation \eqref{matterConservation},
\begin{equation}
\label{GSeta}
\eta=\frac{\gamma\rho v_\text{p}}{B_\text{p}},
\end{equation}
energy conservation \eqref{energyConservation},
\begin{equation}
\label{GSE}
\mathcal{E}=\gamma h\eta-\frac{\Omega_\text{F}I}{2\pi},
\end{equation}
and momentum conservation \eqref{momentumConservation},
\begin{equation}
\label{GSL}
\mathcal{L}=\gamma h\eta r v_\phi-\frac{I}{2\pi},
\end{equation}
where $\Omega_\text{F}=(v_\phi-v_\text{p}B_\phi/B_\text{p})/r$ is the so-called Ferraro isorotation frequency, which is also conserved along the magnetic tube, $v_\text{p}$ and $B_\text{p}$ are the poloidal components of the velocity and magnetic field, and $I$ is the electric current in the tube.

The intensity of radio emission increases with the number of emitting particles, so the radio-emitting areas may simply reflect the areas with an active dense plasma. In this case, the observed three-peaked transverse radio profile (fig.~2 in \cite{Hada2017}) can directly show up the intrinsic structure of the M87 jet and be naturally interpreted thus: the jet as a whole can represent a pinch-like inner jet that is placed in an outer jet, and both jets are coaxial. The inner jet is a solid plasma cylinder of a radius $r_0$, the outer jet is a hollow plasma cylinder of an inner radius $r_1>r_0$ and thickness $d$, so that the radius of the whole jet is $R=r_1+d$.

Equation \eqref{momentumConservation2} implies the following condition at an interface between plasmas with different pressures and electromagnetic fields in the absence of singular density distribution at the interface:
\begin{equation}
\label{boundaryRelation}
\Delta\biggl(p+\frac{B_z^2+B_\phi^2-E_r^2}{8\pi}\biggr)=0,
\end{equation}
where $\Delta$ denotes the difference between the quantities on different sides from the interface.

The basic equations allow singular charge and current densities, such as current sheets, and if a jump in the electromagnetic fields takes place as one passes through the outer boundary of the inner jet or the inner or outer boundary of the outer jet, the corresponding surface charges and currents are non-zero and can readily be found in the standard way. However, the absence of discontinuities and surface charges and currents is beneficial to proper numerical simulations of jets \citep{GourgouliatosEtal2012,KimEtal2017}. On this basis, we may consider surface charges and currents as a potential source of instability and assume their absence. This implies zero jump in electromagnetic fields and, hence, in pressure on the boundaries,
\begin{equation}
\label{boundaryContinuity}
\Delta p=\Delta E_r=\Delta B_\phi=\Delta B_z=0.
\end{equation}
Thus, the electromagnetic field and pressure are continuous everywhere in the case of zero surface charges and currents.

The inner jet bears a total axial electric current $I$ and charge per unit length $Q$. Both the jets may rotate and some toroidal currents are allowed in the plasma, so the electromagnetic field between the jets is
\begin{equation}
\label{electricFieldBetweenJets}
E_r=\frac{2Q}{r},\text{ }B_\phi=\frac{2I}{r},\text{ }B_z=\frac{\alpha}{r}+\beta,\text{ }r_0<r<r_1,
\end{equation}
where $\alpha=(B_{0z}-B_{1z})/(r_0^{-1}-r_1^{-1})$, $\beta=(B_{1z}r_1-B_{0z}r_0)/(r_1-r_0)$, and $B_{0z}$ and $B_{1z}$ are generally different axial magnetic fields on the outer and inner boundaries of the inner and outer jets, respectively. The radial dependence of $B_z$ continuously connects the values of the magnetic fields on the boundaries and corresponds to a concentration of toroidal currents, which can provide a higher magnetization in the centre, towards the inner jet. The case of $\alpha=0$ corresponds to the absence of toroidal currents and a uniform magnetization of the plasma between the jets, $B_z=B_{0z}=B_{1z}$.

We will further consider the case of total charge neutrality and complete current closure, which means that in the outer jet the linear charge density  is $-Q$ and the absolute value of the axial current is $I$, while its direction is opposite to the direction of the axial current in the inner jet. We also assume the absence of any external charges and currents. The total external field then vanishes, $E_r=B_\phi=B_z=0$ for $r>R$.

In principle, the fields outside could have the same structure as those between the jets, so that the external field could be determined by some non-zero charge and current of the jet as a whole, while the axial magnetic field could be generated by some toroidal currents placed far away from the jet. In this case, the energy of the external electromagnetic field goes to infinity, but this fact itself cannot exclude the possibility of non-zero external field as it is largely an artefact of an idealized cylindrical consideration, seeing that the same situation is with an ordinary straight wire with current. Meanwhile, the absence of external field corresponds to full concentration of electromagnetic energy in the jet (which is in a certain sense `energetically profitable' when the internal field remains unchanged) and, as we will discuss below, favours the jet stability.

The quantities $Q$, $I$ and $B_z$ are not independent. We may relate the fields \eqref{electricFieldBetweenJets} by equation \eqref{electricField} to the toroidal and axial velocities $v_\phi$ and $v_z$ of the intermediate plasma:
\begin{equation}
\label{chargeCurrentRelation}
Q=v_zI-\frac{r v_\phi B_z}{2}.
\end{equation}
We may go to the outer boundary of the inner jet or to the inner boundary of the outer jet and relate the above three quantities to the corresponding boundary velocities $v_{0\phi}$ and $v_{0z}$ at $r=r_0$ or $v_{1\phi}$ and $v_{1z}$ at $r=r_1$. On the one hand, the axial magnetic field and velocities of the outer jet are related to the charge and current of the inner jet and, on the other hand, the axial magnetic field of the inner jet is determined by the toroidal current in the intermediate plasma and outer jet, so the jets are not independent but electromagnetically connected.

Equation \eqref{chargeCurrentRelation} implies the relation $v_{0z}-v_{1z}=(r_0 v_{0\phi}B_{0z}-r_1 v_{1\phi}B_{1z})/2I$. When the axial velocities and magnetic fields coincide, $v_{0z}=v_{1z}$ and $B_{0z}=B_{1z}$, the ratio of the azimuthal velocities for the inner and outer jets is the inverse of the ratio for the corresponding radii, $v_{1\phi}/v_{0\phi}=r_0/r_1<1$. In this case, the outer jet rotates slower than the inner and the ratio of angular velocities of the boundaries is $\Omega_1/\Omega_0=(r_0/r_1)^2<1$.

Now notice that for the electromagnetic fields of the form \eqref{electricFieldBetweenJets} we have $(B_\phi^2-E_r^2)'=-2(B_\phi^2-E_r^2)/r$. Substituting these in the momentum conservation law \eqref{momentumConservation2} yields a balance of the relativistic centrifugal force  and the gradient of hydrodynamic plus axial magnetic pressure,
\begin{equation}
\label{intermediateMomentumConservation}
\biggl(p+\frac{B_z^2}{8\pi}\biggr)'=\gamma^2 \rho h\frac{v_\phi^2}{r}.
\end{equation}
For a constant axial magnetic field, we would have a hydrodynamic balance of the pressure gradient and centrifugal force,
\begin{equation}
\label{hydrodynamicMomentumConservation}
p'=\gamma^2 \rho h\frac{v_\phi^2}{r}.
\end{equation}
Since the total momentum conservation in fact signifies the balance of hydrodynamic and electromagnetic forces, the vanishing of electromagnetic fields from the force balance means that the fields so chosen generate zero Lorentz forces and hence the hydrodynamic forces have to balance themselves. Differently speaking, the plasma can bear significant electromagnetic fields but nevertheless flow in a purely hydrodynamic way.

We expect that the plasma density between the jets is significantly lower than in the jets, and hence neglect the centrifugal term in equation~\eqref{intermediateMomentumConservation},
\begin{equation}
\label{intermediatePressureBalance}
p+\frac{B^2_z}{8\pi}=\text{const},\text{ }r_0<r<r_1.
\end{equation}

Using equations \eqref{electricField} and \eqref{electricFieldBetweenJets} and assuming a parabolic longitudinal velocity for the plasma between the jets that coincides with that for the outer and inner boundaries of the inner and outer jets,
\begin{equation}
\label{parabolicAxialVelocityBetweenJets}
v_z=\kappa-\lambda r^2,
\end{equation}
where $\kappa=(v_{0z}r_1^2-v_{1z}r_0^2)/(r_1^2-r_0^2)$ and $\lambda=(v_{0z}-v_{1z})/(r_1^2-r_0^2)$, leads us to the toroidal velocity distribution
\begin{equation}
\label{toroidalVelocityBetweenJets}
v_\phi=\frac{\xi-\zeta r^2}{\alpha+\beta r},
\end{equation}
where $\xi=2(\kappa I-Q)$ and $\zeta=2\lambda I$.

Interestingly, for a constant axial magnetic field, when the plasma moves in a hydrodynamic way, equation \eqref{toroidalVelocityBetweenJets} coincides with that for a viscous Couette flow between rotating cylinders, $v_\phi=\mu/r-\nu r$ with $\mu=\xi/\beta$ and $\nu=\zeta/\beta$, while equation \eqref{parabolicAxialVelocityBetweenJets} coincides with a viscous Poiseuille flow in a pipe when $r_0\ll r_1$ \citep{LandauLifshitz1987}, and viscosity might take part in the formation of the described velocity distributions. When, in addition, we have a constant axial velocity between the jets, the toroidal velocity becomes $v_\phi=\mu/r$ with $\mu=r_0 v_{0\phi}=r_1 v_{1\phi}$.

After the study of the fields and plasmas between the jets, let us devote our attention to the jets themselves. The simplest case for the inner jet corresponds to a uniform distribution of the electric charge and axial current, and equations \eqref{charge} and \eqref{current} give the resulting electromagnetic fields
\begin{equation}
\label{innerJetFields}
E_r=\frac{2Q r}{r_0^2},\text{ }B_\phi=\frac{2I r}{r_0^2},\text{ }B_z=B_{0z}.
\end{equation}
We have assumed here that the axial magnetic field is constant, which implies its generation by toroidal currents outside the inner jet. Assuming a constant axial velocity $v_{0z}$ throughout the inner jet, we get from equations \eqref{electricField}, \eqref{chargeCurrentRelation} and \eqref{innerJetFields} a rigid-like radial distribution of the rotational velocity,
\begin{equation}
\label{innerJetToroidalVelocity}
v_\phi=\frac{v_{0\phi}r}{r_0},
\end{equation}
which corresponds to a constant angular velocity $\Omega_0=v_{0\phi}/r_0$.

When the density~$\rho_0\approx\text{const}$, relativistic enthalpy $h_0\approx1$ and Lorentz factor $\gamma_0\approx\text{const}$, we get from the full momentum conservation law~\eqref{momentumConservation2} a parabolic pressure profile
\begin{equation}
\label{innerJetPressure}
p=p_0-\biggl[\frac{I^2-Q^2}{\pi r_0^2}-\frac{\rho_0(\gamma_0 v_{0\phi})^2}{2}\biggr]\biggl(\frac{r}{r_0}\biggr)^2,\text{ }r<r_0,
\end{equation}
where $p_0$ is the pressure on the jet axis.

It remains to consider the distribution of electromagnetic fields and pressure in the outer jet. The external field is zero, so it is necessary to provide for $E_r$, $B_\phi$ and $B_z$ some transition from their values at $r=r_1$ to zeroes at $r=R$. Let us assume that the width of the outer jet is sufficiently small compared with the width of the whole jet, $d\ll R$. This condition suggests that the above field transition is sufficiently sharp in the sense that $E'_r\sim E_r/d\gg E_r/R$ and $B'_\phi\sim B_\phi/d\gg B_\phi/R$, which gives the inequality $(B_\phi^2-E_r^2)'/8\pi\gg(B_\phi^2-E_r^2)/4\pi r$ and thereby allows us to consider a simplified momentum conservation law
\begin{equation}
\label{outerJetMomentumConservation}
\biggl(p+\frac{B_z^2+B_\phi^2-E_r^2}{8\pi}\biggr)'=\gamma^2 \rho h\frac{v_\phi^2}{r}.
\end{equation}
The simplest linear field transition
\begin{equation}
\label{fieldTransition}
E_r=\frac{2Q}{r_1}\frac{x}{d},\text{ }B_\phi=\frac{2I}{r_1}\frac{x}{d},\text{ }B_z=B_{1z}\frac{x}{d},
\end{equation}
where $x=R-r$, gives a constant toroidal velocity $v_\phi=v_{1\phi}$ for a constant axial velocity $v_z=v_{1z}$.

Thus, the momentum conservation law is analogous to that taking place between the jets in the case of a uniform magnetization, equation \eqref{hydrodynamicMomentumConservation}, with the difference that the ordinary pressure is replaced by the quantity $p+(B^2-E^2)/8\pi$, where the second term may be interpreted as the magnetic pressure in the comoving reference frame. In the laboratory frame, the term $B^2$ is responsible for the magnetic pressure while $-E^2$ for the radial tension of the electric field lines. We finally get the pressure
\begin{equation}
\label{outerJetPressure}
p=p_\text{ex}-\frac{B_z^2+B_\phi^2-E_r^2}{8\pi}-\rho_1(\gamma_1 v_{1\phi})^2\ln\frac Rr,\text{ }r_1<r<R,
\end{equation}
where $\rho_1$ is the plasma density in the outer jet, $\gamma_1$ is the Lorentz factor, $p_\text{ex}$ is the external pressure and the electromagnetic fields are given by equation~\eqref{fieldTransition}. The external medium is subject to equation \eqref{hydrodynamicMomentumConservation} and when it does not rotate, we have some uniform external pressure $p=p_\text{ex}$.

\section{Discussion}

The whole jet-in-jet structure can be explained by the existence of a similar initial density distribution for the flowing plasma near the AGN central engine that plays the role of boundary conditions. In this case, the new radio observations of the innermost region in M87 \citep{Hada2017} can be evidence for simultaneous activity of two different mechanisms of jet launching, one giving a plasma for the inner jet and relating to the central black hole and the other giving a plasma for the outer jet and relating to the accretion disc. In other words, the Blandford-Znajek and Blandford-Payne launching mechanisms can operate at the same time.

In general, the basic equations describing equilibrium and dynamics of jets admit infinitely many possible configurations of gas or plasma flows and concomitant electromagnetic fields, and the fact that we go on observing the same jet-in-jet configuration relatively far from the AGN indicates sufficient stability of the whole jet: the inner and outer jets are straight, gradually widen with distance from the central engine, and their coaxiality is not seen to be violated. How to qualitatively realize the possibility of stability? The electromagnetic field is an interlayer that maintains jet separation. The radial electric field attracts and the toroidal magnetic field repulses the jets, which corresponds to attraction of opposite charges and repulsion of opposite axial electric currents. If the M87 radio data indeed show that we deal with the jet-in-jet structure, then the net effect of the electromagnetic field should correspond to repulsion in order to provide mutual alignment and coaxiality of the jets. This fact requires that the jet necessarily bear significant axial currents because the electric field alone results in an instability with respect to displacement of the central jet from the axis of the outer jet.

Note that if the internal magnetic field were absent, charge separation between the inner and outer jets would vanish due to neutralization of the radial electric field as a result of charge redistribution. The non-zero magnetic field plays a significant role in preventing such redistribution. In the first approximation, a charged particle moves along the magnetic field and simultaneously drifts in the perpendicular direction because of the crossed electric and magnetic fields, so that $\mathbf{v}=v_\parallel\mathbf{B}/B+\mathbf{E}\times\mathbf{B}/B^2$. Since the magnetic field lines are helices while the electric field is radial, the total motion of the charged particle is also helical though, in general, not coincident with a magnetic field line. We see that the electric drift prevents motion of charged particles in the radial direction and thereby maintains charge separation.

The current should stabilize the inner jet position, and the existence of the outer jet plays significant role not only as a way for the inverse current but also as an annular conducting wall that contains the internal toroidal magnetic field and does not allow it to go outside, thereby providing backward elastic forces due to an increase in the denseness of the magnetic field lines when the inner jet is displaced from the axis. Importantly, the longitudinal magnetic field also contributes to stabilization of the outer jet. The jet is not rigid, and if the inner jet will be of a curved form, the stabilizing effect of the outer jet will vanish if this jet takes the same form, seeing that the toroidal magnetic field will then not deform. The axial magnetic field helps the outer jet to maintain rigidity, thus playing the role of a resilient backbone: any bending in the outer jet results in the appearance of a returning force due to a string-like tension of the curved axial field. The same stabilizing effect of the axial field also takes place directly for the inner jet.

In the above stabilization the complete current closure, together with charge neutrality, is very significant because then we do not have a toroidal magnetic field in the external medium and the outside force is determined by the external pressure only. Were the current closure not complete, sufficiently bending the whole get would lead to increasing the denseness of toroidal magnetic field lines on the concave side and give a force that could potentially overcome the stabilizing force of the internal axial field and cause the kink instability of the whole jet. The absence of the external toroidal field and the existence of the internal longitudinal field is also a beneficial factor for stabilization of the sausage instability as we do not have a provocative accumulation of toroidal magnetic field lines around the waist formed in all-round squeezing the jet and at the same time we have a favourable increase in the internal magnetic pressure that resists such squeezing.

This circumstance might explain why astrophysical jets are more stable and lengthy than their laboratory hydrodynamic counterparts: even if the whole jet does not generate external electromagnetic fields and then in this aspect is formally similar to a hydrodynamic one as seen from outside, the existence of an internal electromagnetic structure characteristic of the jet in jet and absent in a hydrodynamic jet gives an extra stabilizing effect.

What provides the current closure and the currents themselves? The jet as a whole is connected to the AGN central engine, a central supermassive black hole with a surrounding accretion disc. Under the condition of infinite conductivity the existence of a vertical magnetic field in the rotating disc results in the appearance of a radial electric field across the disc, so an induced potential difference between the inner and outer jets may be the source maintaining the circulating electric current in the whole jet. This effect is similar to the well-known generation of voltage by a unipolar inductor (Faraday's disc) \citep{Shatskii2003,Okamoto2015}.

This conclusion is consistent with the existing observations. \cite{Hada2017} has been observed that the width of the inner bright part is $r_0\propto z^{0.5}$ while that of the whole jet is $R\propto z^{0.56}$, where $z$ denotes the longitudinal distance. The voltage between the inner and outer jets is the same and does not depend on $z$. It is readily estimated from equation \eqref{electricFieldBetweenJets} as
\begin{equation}
\label{voltage}
V\sim2Q\ln\frac{r_1}{r_0},
\end{equation}
and for constant $V$ and $Q$ we have $r_1\propto r_0$, which implies the same dependence of width on distance for both the jets irrespective of the nature of gradual widening when $R\sim r_1$. Some difference might be assigned to a finite thickness of the outer jet.

The existence of charge separation can also be understood from the Grad-Shafranov approach. We can rewrite equation \eqref{chargeCurrentRelation} as
\begin{equation}
\label{QWithIsorotation}
Q=-\Omega_\text{F} \Psi_\text{eff},
\end{equation}
where $\Omega_\text{F}=(v_\phi-v_z B_\phi/B_z)/r$ is the Ferraro isorotation frequency and $\Psi_\text{eff}=B_z\pi r^2/2\pi$ is an effective normalized magnetic flux in the jet, which coincides with an actual normalized magnetic flux in the case of a constant axial magnetic field. Since we assume a constant field in the inner jet, we have $Q=-\Omega_{\text{F}0}\Psi_0$, where $\Omega_{\text{F}0}$ and $\Psi_0$ is the isorotation frequency and actual flux corresponding to the inner jet. Both the quantities are conserved along the magnetic tube, which may change its radius with the distance from the base, and then can be taken near the central engine. If we have a non-zero magnetic flux and rotation at the base, then the appearance of a charge is necessary and may be considered another reflection of the unipolar induction.

Finally, let us estimate in the framework of the developed jet-in-jet model physical quantities related to the M87 jet. The total intensity $L$ is related to the power released in the jet via operation of the unipolar battery,
\begin{equation}
\label{totalIntensityViaVoltageAndCurrent}
L\sim V I.
\end{equation}
Using equation \eqref{voltage} and estimating $Q\sim I$ (we have considered relativistic motion with $v_z\sim1$ and dropped the rightmost term in equation~\eqref{chargeCurrentRelation}, the validity of which we will check below), we get an estimate for the jet charge and current
\begin{equation}
\label{chargeEstimate}
Q\sim I\sim \sqrt{\frac{L}{2\ln(r_1/r_0)}}.
\end{equation}

The transverse sizes of the inner and outer jets are estimated from the radio intensity slice at $15$~marcsec (projected) from the core presented by \cite{Hada2017} and from his statement that $R/r_0\sim8$. The inner radius of the outer jet is estimated as the semidistance between the peaks as the total length of segments of the line of sight lying in the emitting parts of the hollow cylinder is maximum when the line is tangent to the inner cylinder $r=r_1$. Adopting $1\text{ marcsec}\sim140\text{ }R_\text{Sch}\sim2.5\times10^{17}$~cm, we have the radius of the inner jet,
\begin{equation}
\label{M87InnerJetRadius}
r_{0\,\text{M87}}\sim8\times10^{16}\text{ cm},
\end{equation}
and the inner and outer radii of the outer jet,
\begin{equation}
\label{M87OuterJetRadius}
r_{1\,\text{M87}}\sim4\times10^{17}\text{ cm},\text{ }R_\text{M87}\sim6\times10^{17}\text{ cm}.
\end{equation}

Since the intensity of the M87 jet is estimated from observations as \citep{BroderickEtal2015}
\begin{equation}
\label{M87Intensity}
L_\text{M87}\sim10^{44}\text{ erg s}^{-1}
\end{equation}
(with a factor of few uncertainty), we obtain the linear charge density and the electric current,
\begin{equation}
\label{M87ChargeCurrent}
Q_\text{M87}\sim10^7\text{ C cm}^{-1},\text{ }I_\text{M87}\sim3\times10^{17}\text{ A}.
\end{equation}
The voltage between the inner and outer jets and the jet resistance become
\begin{equation}
\label{M87VoltageResistance}
V_\text{M87}\sim3\times10^{19}\text{ V},\text{ }R^\Omega_\text{M87}\sim100\text{ }\Omega.
\end{equation}
The minimum volume particle densities (in the comoving frame; we adopt the Lorentz factor $\gamma\sim3$ from the observations of \cite{MertensEtal2016}) in the inner and outer jets that are necessary for providing the above charges and currents are
\begin{equation}
\label{M87nE}
n_{\text{e}0\,\text{M87}}\sim10^{-9}\text{ cm}^{-3},\text{ }n_{\text{e}1\,\text{M87}}\sim3\times10^{-11}\text{ cm}^{-3}
\end{equation}
and, as we will see, are much less than the actual particle densities.

Adopting the Kerr parameter $a\sim0.6$, \cite{MertensEtal2016} estimate the isorotation frequency $2.75\times10^{-6}\text{ s}^{-1}$ corresponding to the M87 black hole and conclude that the Blandford-Payne launching mechanism is more appropriate because observations of the jet rotation give the isorotation frequency $1.1\times10^{-6}\text{ s}^{-1}$, which corresponds to the Keplerian radius and velocity
\begin{equation}
\label{M87rBase}
R^\text{base}_\text{M87}\sim10^{16}\text{ cm},\text{ }v^\text{base}_{\phi\,\text{M87}}\sim10^{10}\text{ cm s}^{-1}.
\end{equation}
Since in fact we have both mechanisms, we assign the first frequency to the inner jet and the second to the outer jet,
\begin{equation}
\label{M87OmegaF}
\Omega_{\text{F}0\,\text{M87}}\sim3\times10^{-6}\text{ s}^{-1},\text{ }\Omega_{\text{F}1\,\text{M87}}\sim10^{-6}\text{ s}^{-1}.
\end{equation}
The actual angular velocities of the jets can be estimated from conservation of the quantity $l=\gamma h(1-\Omega_\text{F}rv_\phi)$ along the magnetic tube, which follows from a combination of the integrals \eqref{GSeta}--\eqref{GSL}. Putting $l\approx1$ and $h\approx1$, we arrive at
\begin{equation}
\label{M87OmegaGeneral}
\Omega\sim\frac{1-\gamma^{-1}}{\Omega_\text{F}r^2}
\end{equation}
and get the angular velocities
\begin{equation}
\label{M87Omega}
\Omega_{0\,\text{M87}}\sim4\times10^{-8}\text{ s}^{-1},\text{ }\Omega_{1\,\text{M87}}\sim3\times10^{-9}\text{ s}^{-1}
\end{equation}
and ordinary toroidal velocities
\begin{equation}
\label{M87Vphi}
v_{0\phi\,\text{M87}}\sim3\times10^{9}\text{ cm s}^{-1},\text{ }v_{1\phi\,\text{M87}}\sim10^{9}\text{ cm s}^{-1}.
\end{equation}

Equation \eqref{QWithIsorotation} allows us to express the axial magnetic field via charge, radius and isorotation frequency,
\begin{equation}
\label{BzGeneral}
B_z\sim\frac{2Q}{\Omega_\text{F}r^2},
\end{equation}
and thereby find the axial magnetic fields in the inner and outer jets,
\begin{equation}
\label{M87Bz}
B_{0z\,\text{M87}}\sim0.1\text{ G},\text{ }B_{1z\,\text{M87}}\sim0.01\text{ G}.
\end{equation}
We see that the inner jet is more magnetized than the outer, which can argue that the inner jet forms and collimates due to an effect analogous to the pinch effect: it can be squeezed by a toroidal magnetic field of the axial current (if we drop for clarity the inverse effect of the electric field), and a frozen-in axial magnetic field amplifies during such squeezing. Another possibility is a non-uniform magnetization in the disc, which can form due to some toroidal currents or accumulation of an accreting matter with a frozen-in field near the centre.

The other components of the electromagnetic field are calculated with equation~\eqref{electricFieldBetweenJets}. The radial electric field
\begin{equation}
\label{M87Er}
E_{0r\,\text{M87}}\sim240\text{ V cm}^{-1},\text{ }E_{1r\,\text{M87}}\sim50\text{ V cm}^{-1}
\end{equation}
and toroidal magnetic field
\begin{equation}
\label{M87Bphi}
B_{0\phi\,\text{M87}}\sim0.8\text{ G},\text{ }B_{1\phi\,\text{M87}}\sim0.16\text{ G}
\end{equation}
for the inner and outer jets exceed the above axial magnetic field. We can also verify here that the rightmost term in equation \eqref{chargeCurrentRelation} can be dropped in our estimations: its ratio to the other term, $v_\phi/(B_\phi/B_z)$, is small because $B_\phi\gg B_z$ and $v_\phi\ll1$.

Let us estimate the magnetic flux in the inner jet,
\begin{equation}
\label{M87Phi0}
\Phi_{0\,\text{M87}}=\pi r_0^2 B_{0z}\sim2\times10^{33}\text{ G cm}^{2},
\end{equation}
in the outer jet,
\begin{equation}
\label{M87Phi1}
\Phi_{1\,\text{M87}}=\pi d\biggl(R-\frac23\,d\biggr)B_{1z}\sim4\times10^{33}\text{ G cm}^{2},
\end{equation}
and between the jets,
\begin{equation}
\label{M87PhiIn}
\Phi_{\text{in}\,\text{M87}}=\pi(B_{0z}r_0+B_{1z}r_1)(r_1-r_0)\sim1.3\times10^{34}\text{ G cm}^{2}.
\end{equation}
The total magnetic flux through the jet becomes
\begin{equation}
\label{M87Phi}
\Phi_\text{M87}=\Phi_0+\Phi_1+\Phi_\text{in}\sim2\times10^{34}\text{ G cm}^{2}
\end{equation}
and allows us to find the magnetic field at the jet base under the condition of a uniform magnetization,
\begin{equation}
\label{M87Bbase}
B^\text{base}_\text{M87}\sim\frac{\Phi_\text{M87}}{\pi(R^\text{base}_\text{M87})^2}\sim80\text{ G}.
\end{equation}
Thus, the proposed magnetohydrodynamic model does not require the frequently discussed $B\sim10^3-10^4$~G for explanation of the observed jet luminosity. This result is consistent with the estimate $50\text{ G}<B<124$~G for the magnetic field at the base of the M87 jet derived from the observations of the radio core at $230$~GHz by applying the standard theory of synchrotron radiation \citep{KinoEtal2015}.

The mass density in the disc is estimated from the balance of the magnetic and ram pressures,
\begin{equation}
\label{M87rhoDisc}
\rho^\text{disc}_\text{M87}\sim\frac1{8\pi}\biggl(\frac{B^\text{base}_\text{M87}}{v^\text{base}_{\phi\,\text{M87}}}\biggr)^2\sim3\times10^{-18}\text{ g cm}^{-3},
\end{equation}
which corresponds to the proton number density
\begin{equation}
\label{M87nDisc}
n^\text{disc}_\text{M87}\sim2\times10^6\text{ cm}^{-3}.
\end{equation}

The mass density in the inner jet is estimated from equation \eqref{innerJetPressure} by assuming that the inner jet carries so much plasma as it can still carry, which corresponds to zero square bracket for zero pressure at the interface:
\begin{equation}
\label{M87rho0}
\rho_{0\,\text{M87}}\sim\frac2\pi\biggl(\frac{I}{\gamma_0^2 r_0 v_{0\phi}}\biggr)^2\sim2\times10^{-22}\text{ g cm}^{-3}.
\end{equation}
In the outer jet, we have to use another approach because we do not know a priori the pressure of the external medium that enters equation~\eqref{outerJetPressure}. Assuming that at the base $v_\text{p}\sim v_\phi$ and using conservation of $\eta$ \eqref{GSeta}, we get for $\rho^\text{base}_\text{M87}\sim\rho^\text{disc}_\text{M87}$
\begin{equation}
\label{M87rho1}
\rho_{1\,\text{M87}}\sim\frac{\rho^\text{disc}_\text{M87}v^\text{base}_{\phi\,\text{M87}}}{\gamma_1}\frac{B_{1z\,\text{M87}}}{B^\text{base}_\text{M87}}\sim4\times10^{-23}\text{ g cm}^{-3}.
\end{equation}
We then have the corresponding mass fluxes
\begin{equation}
\label{M87massFluxes}
\dot{M}_{0\,\text{M87}}\sim5\times10^{-3}\text{ M}_\odot\text{ yr}^{-1},\text{ }\dot{M}_{1\,\text{M87}}\sim0.05\text{ M}_\odot\text{ yr}^{-1},
\end{equation}
proton particle densities
\begin{equation}
\label{M87n}
n_{0\,\text{M87}}\sim100\text{ cm}^{-3},\text{ }n_{1\,\text{M87}}\sim30\text{ cm}^{-3},
\end{equation}
and multiplicities $\lambda=n/n_\text{e}$,
\begin{equation}
\label{M87lambda}
\lambda_{0\,\text{M87}}\sim10^{11},\text{ }\lambda_{1\,\text{M87}}\sim10^{12},
\end{equation}
for the inner and outer jets. The total mass flux through the jet,
\begin{equation}
\label{M87totalMassFlux}
\dot{M}_\text{M87}\sim0.05\text{ M}_\odot\text{ yr}^{-1},
\end{equation}
is very large, $\dot{M}_\text{M87}\sim30L_\text{M87}$, and comparable to the measured Bondi accretion rate $0.1-0.2\text{ M}_\odot\text{ yr}^{-1}$ across the Bondi radius of M87 at $0.12-0.22$~kpc \citep{RussellEtal2015}. This circumstance might favour a scenario when almost all initial accretion flow far from the jet goes to the outer jet so that the jet can substantially suppress accretion on to the black hole. Note that a limit of $10^{-3}\text{ M}_\odot\text{ yr}^{-1}$ on the accretion flux is expected from the Faraday rotation measurements of \cite{KuoEtal2014}, but the latest modelling of non-VLBI millimetre data admits much higher accretion rates \citep{MoscibrodzkaEtal2017}.

The external pressure is estimated from equations \eqref{intermediatePressureBalance} and \eqref{outerJetPressure} by adopting zero pressure ar $r=r_0$,
\begin{align}
\label{M87pEx}
p_{\text{ex}\,\text{M87}}&\sim\frac{B_{0z}^2}{8\pi}+\frac1{2\pi}\biggl(\frac{I}{\gamma_1 r_1}\biggr)^2+\rho_1(\gamma_1v_{1\phi})^2\ln\frac R{r_1}
\nonumber\\
&\sim10^{-3}\text{ dyn cm}^{-2},
\end{align}
and implies the jet temperature ($k_\text{B}=1$)
\begin{equation}
\label{M87Temp}
T_\text{M87}\sim\frac{p_{\text{ex}}}{2 n_1}\sim10^{11}\text{ K}.
\end{equation}
This temperature corresponds to non-relativistic protons, $T\sim0.01m_\text{p}$, and relativistic electrons, $T\sim20m_\text{e}$. Temperatures of the same order figured earlier in general relativistic magnetohydrodynamic simulations of the M87 jet \citep{MoscibrodzkaEtal2016,MoscibrodzkaEtal2017}. Interestingly, the temperature slightly exceeds or is comparable to the brightness temperatures $10^{10}-10^{11}$~K observed at smaller distances from the base \citep{KimM87Etal2016}.

\section{Conclusions}

In this paper, new high-resolution VLBI radio observations of the M87 jet at 15 GHz have been considered. A suggestion has been put forward that an unusual three-peaked structure of the transverse radio profile with a previously unobserved ultra-narrow central ridge can reflect the actual structure of the jet. A model in the framework of the relativistic ideal magnetohydrodynamics has been developed that the jet as a whole in fact consists of two coaxial embedded jets such that the outer jet is an annular hollow plasma cylinder that contains a narrow inner jet. The rotating relativistic inner and outer jets move in the same direction away from the central engine and have opposite charges and inversely directed electric currents. The charges, currents and axial magnetic fields are related to the velocities of the electromagnetically connected jets. A continuous distribution of electromagnetic fields and pressure without charge and current sheets is obtained and various physical quantities in relation to the existing observations of the M87 jet are calculated. In particular, the inferred magnetic field at the jet base is about $80$~G and appears sufficient to provide the observed jet luminosity.

The existence of the jet-in-jet structure can directly indicate that two different mechanisms of jet launching simultaneously operate in the central engine: one is responsible for the inner jet and related to the central supermassive black hole and the other is responsible for the outer jet and related to the accretion disc. The central engine plays the role of a battery operating as a unipolar inductor and generating voltage between the inner and outer jets, which provides the current closure in the whole jet. Qualitative consideration of the jet stability shows a beneficial role of the electromagnetic interlayer, charge neutrality and closure of the currents generated by the central engine in stabilization of the sausage and kink instabilities, which might explain why astrophysical jets are more stable than laboratory hydrodynamic jets. The constant voltage between the jets explains similarity between the widening laws for the inner and outer jets, which was found in the VLBI observations.

\section*{Acknowledgements}

The work is supported by the Russian Science Foundation, grant no. 16-12-10051.




\bibliographystyle{mnras}
\providecommand{\noopsort}[1]{}\providecommand{\singleletter}[1]{#1}%


\bsp	
\label{lastpage}
\end{document}